\documentstyle[version2,aps]{revtex}
\begin{document}
\draft
\columnsep -.375in
\twocolumn[
\begin{title}
Josephson Current and Proximity Effect  in Luttinger Liquids
\end{title}
\author{
Dmitrii  L.\ Maslov\rlap,$^{(a,b,d)}$\cite{Maslov}
Michael Stone\rlap,$^{(a)}$\cite{Stone}\\
Paul M.\ Goldbart\rlap,$^{(a,b,c)}$\cite{Goldbart}
Daniel Loss$^{(e)}$\cite{Loss_em}}
\begin{instit}
$^{(a)}$Department of Physics,
$^{(b)}$Materials  Research Laboratory and
$^{(c)}$Beckman Institute,\\
University of Illinois at Urbana-Champaign,
Urbana, Illinois  61801, USA;\\
$^{(d)}$Institute for Microelectronics Technology,
Acad.~of Sciences of Russian Federation
Chernogolovka, 142432, Russia;\\
$^{(e)}$Department of Physics,
Simon Fraser University,
Burnaby B.C. V5A 1S6, Canada
\end{instit}
\receipt{\today}
\begin{abstract}
A theory describing a one-dimensional Luttinger liquid in contact with
superconductor is developed.
Boundary conditions for the fermion fields describing Andreev reflection
 at the
contacts  are derived and used to construct a bosonic representation of the
fermions.
The Josephson current through a superconductor/Luttinger liquid/superconductor
junction
is considered for both perfectly and poorly transmitting interfaces. In the
former
case, the Josephson current at low temperatures is found to be essentially
unaffected by
electron-electron interactions. In the latter case, significant renormalization
of the Josephson
current occurs. The profile of the (induced) condensate wavefunction in a
semi-infinite Luttinger liquid
in contact with a superconductor
is shown to decay as a power-law, the exponent  depending on the sign and
strength of the interactions.
In the case of repulsive (attractive) interactions the decay is faster (slower)
than in their
absence. An equivalent method of calculating the Josephson current through a
Luttinger liquid, which  employs the bosonization of the system as a whole
(i.e., superconductor, as well as
Luttinger liquid) is developed and shown to give the results equivalent to
those obtained via  boundary
conditions describing Andreev reflection.
\end{abstract}
]
\narrowtext
\section{Introduction}
\label{sec:intro}
When a normal metal (N) is put  in good electrical contact with a
superconductor (S), superconducting order is induced in the normal
metal over distances far greater than any microscopic lengthscale,
either of the normal metal  or the superconductor. This induced order
leads to a number of remarkable phenomena, such as the Josephson effect
in SNS junctions \cite{REF:JOSEPH} and the induced Meissner effect in SN
bilayers
\cite{REF:Deutscher}, collectively known as \lq\lq proximity
effects\rq\rq\ \cite{REF:Deutscher}. Until very recently, all work on such
effects, both experimental and theoretical, has concentrated on systems
in which N is in the Fermi-liquid (FL) state. It has long been
appreciated theoretically that, in contrast with their
higher-dimensional analogues, (effectively) one-dimensional systems of
interacting electrons are not Fermi liquids. Instead, they exhibit
a number of possible regimes \cite{REF:fn_states:Emery}, among which
the Luttinger liquid (LL) provides a  one-dimensional (1D) metallic
counterpart to the (higher-dimensional) FL state, albeit differing  in
several important respects, most notably in the absence of
single-particle excitations in the low-energy part of the spectrum.
The basic features of LLs have been  understood mainly in
the context of 1D organic charge-transfer and mixed-valence conductors
\cite{REF:reviews}.  In addition, the prediction of
the suppression of the tunneling conductance of LLs \cite{REF:KF1,REF:KF2}
has stimulated the experimental search for
Luttinger liquids in mesoscopic systems, in particular, in the edge-channels of
fractional quantum Hall systems \cite{REF:MILLIK} and in semiconductor
quantum wires \cite{REF:NTT}.

The purpose of this paper is to address the issue of proximity effects
at Luttinger-liquid/super{\-}con{\-}ductor interfaces, including the Josephson
effect in superconductor/Luttinger-liquid/superconductor (SLS)
junction. Our motivation is twofold. First, experimentally, such a
study is relevant in view of the rapid progress in the fabrication of
supercon{\-}ductor/semi{\-}conductor interfaces \cite{REF:exp.review.SSm},
especially those with high inter{\-}face-trans{\-}parency (such as, e.g., the
Nb/InAs interface), and also  in view of the recently reported observations
of LL-like  behavior in
GaAs quantum wires \cite{REF:NTT}.  Thus, the fabrication and
investigation of SLS  systems may reasonably  be anticipated in the
near future. Second, theoretically, we aim to understand the interplay
between  electron-electron interactions and induced  superconducting
order in 1D electronic systems. Furthermore, one of the possible
scenarios of high-temperature superconductivity in oxide materials is
built on the assumption of the LL-like character of the normal
electronic state in these materials \cite{REF:ANDERS}.  The existence
of LLs in dimensions higher than one, however, is not yet established,
in contrast to the 1D case. Thus, a 1D LL in which the
superconductivity is induced via the proximity effect may provide a model
system for
superconductivity in 2D
\cite{REF:fn_LL_pairing}.

Our main results can be formulated as follows: (i)~At low temperatures,
the Josephson
current through an SLS junction having  perfectly transmitting
interfaces has the same phase- and length-dependences as in
the noninteracting case, the only difference being a renormalization
of the effective Fermi velocity. The reason for this is that, using the bosonic
language, the non-dissipative (topological) currents, including the
Josephson current, are carried in LLs by the topological modes of the
boson fields, which are not sensitive to the interactions. At
temperatures above a certain crossover value, interactions lead to the
additional suppression of the Josephson current.
(ii)~The (induced) superconducting condensate wavefunction in a LL in good
electrical contact with a superconductor decays as $x^{-\gamma}$ with
the distance $x$ from the LL/S interface, with $\gamma$ depending on
the strength of the interactions. ($\gamma=1$ corresponds to the
noninteracting case, whereas $\gamma>1$ for repulsive and $\gamma<1$
for attractive interactions.) (iii)~For the case of imperfectly
transmitting interfaces, the renormalization of the interface-transmission
 coefficients (via a mechanism known from studies of LLs
coupled by weak links \cite{REF:KF1,REF:KF2}) is reflected in the
renormalization of
the Josephson current, which gets strongly suppressed in the case of
the repulsive interactions. Along the way, we have also: (iv)~derived
effective boundary conditions describing  Andreev reflection at the
interface with a superconductor, which we have then used as boundary
conditions in the bosonization procedure; (v)~determined the structure
of the topological (Haldane) excitations in an SLS system; and
(vi)~confirmed result~(i) via an alternative approach, in which
bosonization is applied to both the superconducting and normal parts of
the system.

The issue of the Josephson current through a LL has also been studied
in a recent paper by Fazio, Hekking, and Odintsov \cite{REF:Hekking} for
the case of poor  interface-transmittance (see also Ref.~\cite{lutt_and}).  By
using the tunneling
Hamiltonian method, it was found that the Josephson current through an
SLS junction is suppressed compared to the noninteracting
case. The present paper takes a  different approach. This approach
originates  from  work on SNS junctions with perfect
interface-transmittance
\cite{REF:Kulik,REF:Ishii,REF:Svid,REF:Bardeen}, in which the Josephson
current was related to the spectrum of electronic states confined to
the N region by Andreev reflection. Our results concerning poorly
transmitting interfaces agree  with  those of Ref.~\cite{REF:Hekking} (up
to non-universal numerical prefactors, which we have not attempted to
calculate).

The present  paper is organized as follows. In Sec.~\ref{sec:abc} we
derive the boundary conditions for the fermion field operators at the
NS interface in the absence of interactions.  In Sec.~\ref{sec:bos} we develop
a  bosonization
procedure for interacting fermions confined to the normal 1D region of
an SLS system, which makes use of the boundary conditions derived in
Sec.~\ref{sec:abc}. We calculate the Josephson current through an SLS
junction in Sec.~\ref{sec:jos}. In Sec.~\ref{sec:prox}, we analyze the
profile of the condensate amplitude in a semi-infinite LL connected to
a superconductor. Up to this stage,  we will have applied bosonization
only to the normal part of the system, the presence of a superconductor
being implemented as a boundary condition.  An alternative approach, in
which both the normal and the superconducting parts of the system are
bosonized, is presented in Sec.~\ref{sec:ms}.
\section{Andreev boundary conditions}
\label{sec:abc}
\subsection{Andreev reflection at the NS interface: qualitative picture}
\label{sec:aqp}
Electronic excitations in a normal metal having energies smaller than
the superconducting energy gap $\Delta_0$ suffer Andreev reflection at
the interface, i.e., electron-like excitations are reflected as
hole-like excitations, with a Cooper pair being injected into the
superconductor, and vice versa \cite{REF:Andr,REF:fn.normal}.  The
single-particle excitations in S are  mixtures of electron- and
hole-like states with weights determined by the self-consistency
condition. In the bulk of N,  the electron- and hole-like states are
uncorrelated. Near the boundary, however, Andreev reflection mixes the
electron- and hole-like states precisely in the same proportion as they
are mixed in S, which leads to the formation of a condensate,  the
amplitude of which decays into the bulk of N. The decay-length of this
condensate is the length $L_T$ over which superconducting correlations in the
motion of
bulk normal-state electrons  exist. In the case of perfect metals,
$L_T=\hbar v_{\rm F}/T$, where $v_{\rm F}$ is the Fermi velocity and $T$ is the
temperature (we choose units in which $k_{\rm B}=1$).
(The same length determines the thermal disruption
(in the absence of inelastic processes) of mesoscopic phase-coherence, as is
manifested
in the phenomenon of universal conductance fluctuations \cite{REF:ucf}).
For $T\ll \Delta\simeq T_{\rm c}$ (where $T_{\rm c}$ is the critical
temperature of
S), $L_T\gg\xi_{\rm S}$ (where $\xi_{\rm S}\simeq\hbar v_{\rm F}/\Delta_0$ is
the coherence length of S).
In order to describe the influence of the superconductors
on the N region, we now derive effective boundary conditions that
account for the Andreev reflection suffered by the low-energy
components of the fermion fields at the NS interfaces. Our
strategy is as follows: in Sec.~\ref{sec:dabc},  we derive these boundary
conditions for
the case of the non-interacting electron gas in N; then, in Sec.~\ref{sec:bos},
we
implement these boundary conditions into the bosonization scheme for
interacting electrons.
\subsection{Derivation of Andreev boundary conditions}
\label{sec:dabc}
We consider a one-dimensional electronic conductor (i.e., a quantum
wire) of length $L$, adiabatically connected to superconducting leads
(see Fig.~\ref{FIG:fig1}a). We begin by analyzing the ideal case, in which the
single-electron parameters (Fermi velocities, effective masses, etc.)
are the same in the N and S parts of the structure, the only difference
between N and S being the presence of a pairing potential in S.  We
adopt the conventional model
\cite{REF:Kulik,REF:Ishii,REF:Svid,REF:Bardeen} in which the pairing
potentials in the leads  are assumed to be unaffected by the presence of
N.  Although this is a non--self-consistent approximation, it is known to
reflect correctly the aspects of the problem relevant for the present
treatment \cite{REF:Kulik,REF:Ishii,REF:Svid,REF:Bardeen}.

We temporarily replace the real 3D superconducting leads by effective
1D leads.  The  profile of the pair potential is then given by
 (cf.~Fig.~\ref{FIG:fig1}b)
\begin{equation}
\Delta(x)=
\cases{
 \Delta_0 e^{i\chi_1}&for $x\leq 0$;\cr
0&for $0<x<L$;\cr
\Delta_0e^{i\chi_2}&for $x\geq L$.
 }
 \label{eq:delta}
 \end{equation}
 In the Andreev (semi-classical) approximation \cite{REF:Andr},
which is valid
 for $\Delta_0\ll\epsilon_F\equiv\hbar k_{\rm F}v_{\rm F}/2$,
 the  spinor of Bogoliubov amplitudes
 \begin{eqnarray}
&&{\bf w}=\left(
 \begin{array}{l}
 u\\
 v\\
 \end{array}\right)
 \label{eq:nambu}
 \end{eqnarray}
 satisfying the Bogoliubov-de~Gennes equations \cite{REF:deGennes} is
 decomposed into left- and right-moving components,
 \begin{equation}
 {\bf w}=e^{ik_{\rm F}x}{\bf w}_{+}+e^{-ik_{\rm F}x}{\bf w}_{-},
 \label{eq:decomp}\end{equation}
where $k_{\rm F}$ is the Fermi wavevector. The components ${\bf w}_{\pm}$ now
satisfy
the (formally) relativistic (first-order) Bogoliubov-de~Gennes
equations:
 $H^{\pm}_D{\bf w}_{\pm}=\epsilon
 {\bf w}_{\pm}$, with the
Hamiltonians
\begin{eqnarray}
H^{\pm}_D\:  &=& \:
\left(
 \begin{array}{cc}
 \mp  i\hbar v_{\rm F}\partial _x & \Delta(x)\nonumber\\
 \Delta^{*}(x) & \pm i\hbar v_{\rm F}\partial _x\nonumber
 \end{array}\right).
 \label{eq:dirac}
 \end{eqnarray}
 The full solution of these equations is obtained
  \cite{REF:Kulik} by finding the solutions
 in the N and S regions and then matching them at the interfaces.
 (In the semi-classical approximation, only the wavefunctions
 need be continuous.)\thinspace\ The solution in N
 for $\epsilon<\Delta_0$
 can be written as
 \begin{eqnarray}
{\bf w}_{\pm}\:  &=& \:
A_{\pm}\left(
 \begin{array}{l}
 e^{\pm ikx}\\
{\cal R}^{\mp 1}(\epsilon)e^{-i\chi_1}e^{\mp ikx}
 \end{array}\right),
 \label{eq:bdgp}
 \end{eqnarray}
 where
 \begin{equation}
 \label{eq:refl}
 {\cal
 R}(\epsilon)=e^{-i\eta(\epsilon)}\quad{\rm and}\quad\eta(\epsilon)=
 \cos^{-1}(\epsilon/\Delta_0),
 \end{equation}
in which  ${\cal R}$ is the Andreev reflection coefficient,
whose phase is $\eta$.  The quasiparticle momentum
$\hbar k=\epsilon/v_{\rm F}$ satisfies the quantization condition
\begin{equation}
 {\cal R}(\epsilon)^2e^{\pm i(\chi_1-\chi_2)}e^{2ikd}=1,
\label{eq:eigen}\end{equation}
where $\pm$ corresponds to two sets of energy levels \cite{REF:Kulik}.
In Eq.~(\ref{eq:bdgp}), $A_{\pm}$ are overall
normalizations which, without the loss of generality, can be chosen to
be real. Evaluating Eq.~(\ref{eq:bdgp}) at $x=0$ and
$x=L$ and using Eq.~(\ref{eq:eigen}), one can see that at the
boundaries the left and right components of the Bogoliubov amplitudes
satisfy
\begin{equation}
\cases{
v_{\pm}={\cal R}^{\mp 1}e^{-i\chi_1}u_{\pm},&for $x=0$,\cr
v_{\pm}={\cal R}^{\pm 1}e^{-i\chi_2}u_{\pm},&for $x=L$.}
\label{eq:uvbc}
\end{equation}
Equations~(\ref{eq:uvbc}) describe the essence of Andreev reflection:
the electron-like excitations ($u_{\pm}$) are converted into hole-like
excitations ($v_{\pm}$), at the same time
acquiring the phase of the order parameter ($\chi_{1,2})$ together
with the phase-shift of the Andreev reflection coefficient ($\eta$).

In the limit $\epsilon\ll \Delta_0$, the phase shift $\eta\to\pi/2$
\cite{fn:eta},
and the boundary conditions (\ref{eq:uvbc}) become energy-independent.
This enables one to derive from Eqs.~(\ref{eq:uvbc}) the boundary
conditions for the real-space fermion operators $\psi_s(x)$,
where $s={\uparrow},{\downarrow}$ denotes the spin projection. These
field operators are related to the $(u,v)$ amplitudes via the
Bogoliubov transformation \cite{REF:deGennes}
\begin{equation}
\psi_{s}(x)=\sum\left(c_su(x)-sc^{\dagger}_{-s}v^{*}_{s}(x)\right),
\label{eq:bt}\end{equation}
where $c_s$ ($c^{\dagger}_s$) is the fermion annihilation (creation)
operator, the sum runs over all
single-particle quantum numbers, and the variable $s$ takes on the
values $+1 (-1)$ for the $\uparrow (\downarrow)$ spin-projections.  We
decompose $\psi_s(x)$ into the left- and right-movers:
\begin{equation}
\psi_s(x)=e^{ik_{\rm F}x}\psi_{+,s}(x)+e^{-ik_{\rm F}x}\psi_{-,s}(x).
\label{eq:lrdec}\end{equation}
Substituting decompositions (\ref{eq:decomp}) and (\ref{eq:lrdec}) into
Eq.~(\ref{eq:bt}), we obtain the Bogoliubov transformation for
$\psi_{\pm,s}$:
\begin{equation}
\label{eq:bogol}
\psi_{\pm,s}(x)=\sum
\left(c_su_{\pm}(x)-sc_{-s}^{\dagger}v_{\mp}^{*}(x)\right).
\end{equation}
The boundary conditions for the (Pauli) spinors $\psi_{\pm ,s}$ then
follow upon substitution of Eqs.~(\ref{eq:uvbc}) into
Eq.~(\ref{eq:bogol}), and using  $\eta=\pi/2$. After some algebra, we
obtain
\begin{mathletters}
\begin{eqnarray}
\psi_{+,\uparrow}\big\vert_{x=0,L}&=&\mp ie^{i\chi_{1,2}}
\psi_{-,\downarrow}^\dagger\big\vert_{x=0,L}\,\,,\label{eq:lrbc1a}\\
\psi_{+,\downarrow}\big\vert_{x=0,L}&=&\pm ie^{i\chi_{1,2}}
\psi_{-,\uparrow}^{\dagger}\big\vert_{x=0,L}\,\,,
\label{eq:lrbc1b}\end{eqnarray}\end{mathletters}
or, more compactly,
\begin{eqnarray}
\left(
\begin{array}{l}
\psi_{+,\uparrow}
\\
\psi_{+,\downarrow}
\end{array}
\right)\Big\vert_{x=0,L}=\mp ie^{i\chi_{1,2}}\,{\hat T}\left(
\begin{array}{r}
\psi_{-,\uparrow}
\\
\psi_{-,\downarrow}
\end{array}
\right)\Big\vert_{x=0,L}.
\label{eq:lrbc}
\end{eqnarray}
Here, ${\hat T}={\hat g}{\hat C}$ is the time-reversal
operator \cite{REF:deGennes}, with
\begin{eqnarray}
{\hat g}\:  &=& \:i{\hat \sigma}_y=
\left(
 \begin{array}{cc}
 0 & 1\\
 -1 & 0
 \end{array}\right)
 \label{eq:metric}
 \end{eqnarray}
and ${\hat C}$ being the Hermitian conjugation operator.  The presence of
${\hat T}$ in Eq.~(\ref{eq:lrbc}) signals an important property of
Andreev reflection \cite{REF:Andr}:  a reflected excitation is the
time-reversed version of an  incident one.

Further insight into the meaning of the boundary conditions
(\ref{eq:lrbc}) can be obtained by employing the chiral symmetry of
left-right fermion fields $\psi_{\pm,s}$. In what follows, we adopt the
methods of Refs.~\cite{REF:Eggert,REF:Polch}, in which  the chiral
symmetry of boson fields satisfying Dirichlet or  Neumann boundary
conditions was used to derive effective periodic boundary
conditions.  The right (left) field describes the propagation of the (formally)
relativistic fermions to the right (left) with the Fermi velocity.
Consequently, in the Heisenberg representation, the space-time
dependence of these fields is given by
\begin{equation}
\psi_{\pm, s}(x,t)=\psi_{\pm, s}(x\mp v_{\rm F}t).
\label{eq:waves}\end{equation}
Using the boundary conditions (\ref{eq:lrbc1a},\ref{eq:lrbc1b}), one
sees that at  any instant of time the time-dependent left-moving
fermion fields  satisfy
\begin{mathletters}
\begin{eqnarray}
\psi_{-,s}^{\dagger}(v_{\rm F}t)&=&-sie^{-i\chi_1}
\psi_{+,-s}
(-v_{\rm F}t),
\label{eq:bcla}\\
\psi_{-,s}^{\dagger}(L+v_{\rm F}t)&=&s ie^{-i\chi_2}
\psi_{+,-s}(L-v_{\rm F}t).
\label{eq:bclb}
\end{eqnarray}
\end{mathletters}Choosing $t=(L+x)/v_{\rm F}$
 in Eq.~(\ref{eq:bclb}), we obtain
$\psi_{-,s}^{\dagger}(x+2L)=sie^{-i\chi_2}
\psi_{+,-s}(-x)$.
Equation~(\ref{eq:bcla}) gives $\psi_{+,s}(-x)
=-sie^{i\chi_1}
\psi_{-,-s}^{\dagger}(x)$ which, in combination with the previous equation,
leads to
\begin{mathletters}
\begin{eqnarray}
&\psi_{-,s}(x+2L,t)=e^{i\pi\vartheta}\psi_{-,s}(x,t),
\label{eq:pera}\\
&\psi_{+,s}(x,t)=-sie^{i\chi_1}\psi_{-,-s}^{\dagger}(-x,t),
\label{eq:perb}\end{eqnarray}
\end{mathletters}where $\vartheta\equiv 1+(\chi/\pi)$ and $\chi\equiv
\chi_2-\chi_1$.  Thus, we see that the Andreev boundary
 conditions~(\ref{eq:lrbc1a},\ref{eq:lrbc1b}) are equivalent to twisted {\it
periodic}
boundary conditions for $\psi_{-,s}$, Eq.~(\ref{eq:pera}), on an
interval of length twice the length of the original system,
supplemented by the connection between the $\psi_{+,s}$ and
$\psi_{-,s}$ fields following from the chiral symmetry,
Eq.~(\ref{eq:perb}). (Equivalently, the periodic boundary conditions
can be derived for $\psi_{+,s}$ and the chiral symmetry can be used to
obtain $\psi_{-,s}$.)\thinspace\ The problem thus becomes  very similar to one
of fermions on a ring of circumference $2L$, threaded by an
effective Aharonov-Bohm flux $\vartheta/2$, the persistent current
\cite{REF:GI,REF:Kulik_pers,REF:BYL} being the analogue of the Josephson
current.
A detailed treatment  of persistent currents
in Luttinger liquids was given in Ref.~\cite{REF:Loss} (for the case of
spinless electrons), and we shall adopt this treatment in what follows.
There are significant differences between the system of electrons on a
ring and an SLS junction, however: (i)~On a ring, the number of
electrons  is fixed.  Therefore, the system is described by the
canonical ensemble, and is sensitive to the parity of the total number
of fermions \cite{REF:Leggett,REF:Loss}. In the normal part of an SLS
junction, the number of electrons
 is not fixed, so that the system must be described by the grand
canonical ensemble, the chemical potential being maintained in the
superconducting leads. (ii)~On a ring, all four components of the
fermion fields  ($\psi_{\pm,\uparrow,\downarrow}$) satisfy equivalent
twisted periodic boundary conditions. In an  SLS junction, only two of
the four components satisfy twisted periodic boundary conditions,
Eq.~(\ref{eq:pera}); once these components are constructed, the other
two are found using the chiral symmetry, Eq.~(\ref{eq:perb}).

We now discuss the range of validity of the Andreev
boundary conditions. The condition $\epsilon\ll \Delta_0$, which we
needed in order to arrive at Eq.~(\ref{eq:lrbc}), means that our
boundary conditions are capable of describing only
excitations with wavelengths $1/k\gg\hbar v_{\rm
F}/\Delta_0\simeq\xi_{\rm S}$.  Such excitations exists only in \lq\lq
long\rq\rq\ junctions, i.e., $L\gg\xi_{\rm S}$; thus our treatment is
valid only for this case.  On the other hand, as follows from
self-consistent calculations \cite{REF:Deutscher}, the order parameter in
S gets reduced from its bulk value over the scale $\xi_{\rm S}$ near
the boundary, which also affects the excitations in N in the boundary
region of the thickness of $\xi_{\rm S}$. Thus, the model of a
step-like profile  of $\Delta$, Eq.~(\ref{eq:delta}), can adequately
describe only processes taking place in the interior of N (i.e., for
$x$ outside boundary layers of width $\xi_{\rm S}$), where the exact shape of
the
profile of $\Delta$ in S is irrelevant.  The latter
condition can be satisfied only if
$L\gg\xi_{\rm S}$. Therefore, the range of
validity of our boundary conditions is the same as that of the
non--self-consistent model itself. We can also view Eq.~(\ref{eq:lrbc})
as the minimal-model boundary conditions that describe the
time-reversal process associated with Andreev reflection.  Thus, we
now relax the assumption of 1D superconducting leads, and regard
Eqs.~(\ref{eq:lrbc}) as the general boundary conditions satisfied by
the low-energy components of the fermion fields.

\section{Bosonization of Luttinger liquid in contact with superconductors}
\label{sec:bos}
We now turn to the bosonization of an interacting 1D electronic system
\cite{REF:MSB}  in contact with superconductor. We represent the free
fermion fields in the conventional bosonic form:
\begin{equation}
\psi_{\pm,s}(x)=\frac{1}{\sqrt{\alpha L}}\exp\left(\pm
i\sqrt{\pi}\phi_{\pm,s}(x)\right),
\label{eq:mand}
\end{equation}
where $\alpha\rightarrow +0$ is a convergence factor and the chiral
bosons $\phi_{\pm,s}$ are expressed through the density (phase) bosons
$\phi_s (\theta_s)$ via
\begin{equation}
\phi_{\pm,s}(x)=\phi_s(x)\mp\theta_s(x).
\label{eq:chbos}\end{equation}
We construct mode expansions for $\phi_{\pm,s}(x)$ in such a way
that the twisted boundary conditions (\ref{eq:pera}) and the  auxiliary
conditions (\ref{eq:perb}) are satisfied:
\begin{mathletters}
\begin{eqnarray}
\phi_{-,s}(x)&=&\frac{\varphi_{s}}{\sqrt{\pi}}+\sqrt{\pi}(N_s
+\vartheta)
\frac{x}{2L}+{\bar\phi}_s(x),\label{eq:mmin}\\
\phi_{+,s}(x)&=&\frac{\varphi_{-s}}{\sqrt{\pi}}-\sqrt{\pi}(N_{-s}+\vartheta)
\frac{x}{2L}+{\bar \phi_{-s}(-x)}.\label{eq:mplus}
\end{eqnarray}
\end{mathletters}(The additive c-number terms have been omitted in the
expansion for $\phi_{+,s}$.) Here, $\varphi_s$ are zero-mode operators, $N_s$
are operators whose eigenvalues give the winding numbers of the
Haldane (topological) excitations \cite{REF:Haldane}, and ${\bar
\phi}_s(x)$ are the nonzero-mode components of the chiral boson fields,
which are periodic on the interval $(0,2L)$:
\begin{equation}
{\bar \phi_s(x)}=\sum_{k>0}\gamma_k\Big(e^{-ikx}a^{\dagger}_{k,s}+e^{ikx}
a_{k,s}^{{\phantom\dagger}}\Big),
\label{eq:chirper}
\end{equation}
where $k=\pi n/L$ (with $n=1, 2, \dots$),
$\gamma_k=\exp(-\alpha n/2)/\sqrt{kL}$, and $a_{k,s}$
satisfy the canonical commutation relations
$[a_{k,s}^{\phantom{\dagger}},a^{\dagger}_{k',s'}]=\delta_{ss'}\delta_{kk'}$.
In Eqs.~(\ref{eq:mmin},\ref{eq:mplus}), the terms linear in $x$
describe
the topological excitations of the bosonic  system, which do not
conserve the total number of fermions. The eigenvalues of $N_s$ give
the numbers of fermions added to or removed from the Luttinger liquid.
The nonzero-mode components ${\bar \phi_s}$ describe the quantum
fluctuations around the topological excitations. These correspond to
the fluctuations in the fermion density that  conserve the total number
of fermions.

We require that the chiral bosons obey the canonical commutation relations
\cite{REF:Shankar}
\begin{equation}
[\phi_{\pm,s}(x),\partial_{x'}\phi_{\pm, s'}(x')]
=\mp i\delta_{ss'}\sum_{n=-\infty}^{\infty}
\delta(x-x'+2nL),
\label{eq:ccrchir}
\end{equation}
where the summation over $n$ reflects periodicity on the interval
(0,2L). The nonzero-mode components of, e.g.,  expansion
(\ref{eq:mmin}) obey
\begin{eqnarray}
&&[{\bar \phi}_{-,s}(x),\partial_{x'}{\bar \phi}_{-,s'}(x')]\nonumber\\
&&\qquad=\delta_{ss'}\Big\{ix/L+
\sum_{n=-\infty}^{\infty}\delta(x-x'+2nL)\Big\}.
\label{eq:nmccr}\end{eqnarray}
Thus, in order for Eq.~(\ref{eq:ccrchir}) to be satisfied, the
zero-mode operators
 $\varphi_s$ and the winding-number operators $N_s$ must obey
\begin{equation}
[\varphi_s,N_{s'}]=2i\delta_{ss'}.
\label{eq:zmnc}\end{equation}
(The same result can certainly be found by considering the commutation
relations of $\phi_{+,s}$.) {\it A posteriori}, we can also justify the
choice of the coefficients $\gamma_k$ in
Eqs.~(\ref{eq:mmin},\ref{eq:mplus}): they were chosen in such a way
that the commutation relations (\ref{eq:ccrchir}) are satisfied.

Next, we introduce the charge ($\rho$) and spin ($\sigma$) components
of the boson fields:
$\phi_{\rho,\sigma}\equiv (\phi_{\uparrow}\pm\phi_{\downarrow})/\sqrt{2}$ and
$\theta_{\rho/\sigma}\equiv
(\theta_{\uparrow}\pm\theta_{\downarrow})/\sqrt{2}$.
The mode expansions for $\phi_{\mu}$ and $\theta_{\mu}$ (where
$\mu=\rho, \sigma$) follow from the expansions
(\ref{eq:mmin},\ref{eq:mplus}):
\begin{mathletters}
\begin{eqnarray}
\phi_{\rho}(x)&=&\frac{\varphi_{\rho}}{\sqrt{\pi}}+\sum_{k>0}\gamma_k\cos kx
\left(a_{k\rho}^{\dagger}+a_{k\rho}^{{\phantom\dagger}}\right),
\label{eq:phirho}\\
\phi_{\sigma}(x)&=&\sqrt{\frac{\pi}{2}}M\frac{x}{L}+i\sum_{k>0}\gamma_k
\sin kx \left(a_{k\sigma}^{\dagger}-a_{k\sigma}^{{\phantom\dagger}}\right),
\label{eq:phisig}\\
\theta_{\rho}(x)&=&\sqrt{\frac{\pi}{2}}(J+\vartheta)\frac{x}{L}
+i\sum_{k>0}\gamma_k\sin kx
\left(a_{k\rho}^{\dagger}-a_{k\rho}^{{\phantom\dagger}}\right),
\label{eq:thetarho}\\
\theta_{\sigma}(x)&=&\frac{\varphi_{\sigma}}{\sqrt{\pi}}
+\sum_{k>0}\gamma_k\cos kx
\left(a_{k\sigma}^{\dagger}+a_{k\sigma}^{{\phantom\dagger}}\right),
\label{eq:thetasig}
\end{eqnarray}
\end{mathletters}where $\varphi_{\rho/\sigma}
\equiv(\varphi_{\uparrow}\pm\varphi_{\downarrow})/
\sqrt{2}$, $M\equiv (N_{\uparrow}-N_{\downarrow})/2$,
$J\equiv (N_{\uparrow}+N_{\downarrow})/2$, and
$a_{k\rho/\sigma}\equiv(a_{k,\uparrow}\pm a_{k,\downarrow})/\sqrt{2}$.
It is natural that the phase difference of the superconducting
order parameters $\chi$,
which determines  the charge flow between the superconductors, appears
only in the field associated with  the charge
current, i.e., $\theta_{\rho}$.

We now have to determine  the topological constraints imposed on the winding
numbers $N_s$ (and, consequently, on $M$ and $J$). This can be done by
substituting  the expansion, e.g., for $\phi_{-,s}$, Eq.~(\ref{eq:mmin}),
into the bosonization formula (\ref{eq:mand}),  and requiring that
the boundary conditions  for fermions (\ref{eq:pera})  be
satisfied \cite{REF:Loss}. [When disentangling the operators in the
exponent of Eq.~(\ref{eq:mand}), one must recall that $\phi_s$ and $N_s$
do not commute, and use Eq.~(\ref{eq:zmnc}).]\thinspace\
One thus finds that $N_s$ satisfies
\begin{equation}
(-)^{N_s+1}=1,
\label{eq:constr}\end{equation}
i.e., that the eigenvalues of $N_s$ are odd.
(Neglecting the operator-nature of the zero-modes
and the winding numbers would have led to $N_s$ being even.) Consequently,
$J+M$ must be odd. It is convenient to introduce an effective winding-number
 $J'=J+1$, so that  $J+\vartheta=J'+\chi/\pi$ in Eq.~(\ref{eq:thetarho}). Then,
$J'+M$ must be even. Comparing this
constraint with the similar constraint on the topological numbers in the
persistent-current problem \cite{REF:Loss}, we see that our constraint
effectively
corresponds to the case of an {\it odd\/} number of fermions on the
ring, in which case the response of the system to the twist  in boundary
conditions
is diamagnetic, i.e., the free energy is minimal at zero twist. Tracing back
through our calculations, we note that the diamagnetic nature of the
Josephson current is guaranteed by the Andreev phase shift ($\pi/2$),
which ultimately shifts $J$ to $J+1$.
The physical meaning of the topological constraint is quite simple:
the energy of the LL is minimal when the left- and right-moving branches
of the spectrum are populated  symmetrically; changing the total number of
fermions in the LL by an even (odd)
number  results in  excitations
with even (odd) total momentum quanta.

We note that expansions similar to
Eqs.~(\ref{eq:phirho},\ref{eq:phisig},\ref{eq:thetarho},\ref{eq:thetasig})
could have been
obtained by first deriving the boundary conditions directly
for the charge and spin bosons
from the boundary conditions for the fermions (in the same way that the
boundary
conditions for bosons are derived from the Dirichlet boundary conditions for
fermions
in Ref.~\cite{REF:Eggert}), and then constructing expansions satisfying
these boundary conditions. In this way, however, the zero-modes of the
expansions, which are crucial for the topological constraints on the
eigenvalues of $M$ and
$J$, might have
been missed. (We will derive and use the boundary conditions for charge/spin
bosons
 in Sec.~\ref{sec:prox}, when the
topological structure of the boson fields will not be important.)\thinspace\

The bosonized Hamiltonian of the LL is given by
\begin{equation}
{\cal H}=\frac{\hbar}{2}\!\sum_{\mu=\rho,\sigma}\!\int_{-L/2}^{L/2}dx\left\{
\frac{v_{\mu}}{K_{\mu}}(\partial_x \phi_\mu)^2+
v_{\mu}K_{\mu}(\partial_x \theta_\mu)^2\right\}.
\label{eq:llham}\end{equation}
If the LL model originates from the Hubbard model then
$K_{\rho/\sigma}=1/\sqrt{1\pm g}$, where $g\equiv Ua/\pi v_{F}$,
with $U$ being the strength of on-site interactions and
$a$ the microscopic length cut-off (of order
the Fermi wavelength), and $v_{\mu}\equiv
v_{F}/K_{\mu}$. In addition, if the underlying SU(2) symmetry of the
Hubbard model is intact, then $K_\sigma=1$ \cite{REF:Emery}.

\section{Josephson current through a Luttinger liquid}
\label{sec:jos}
One of the most important consequences of  induced coherence in the
N part of an SNS system, is the
Josephson current through it.  This current differs from its counterpart
in tunnel junctions in that the critical current ${\cal J}_c$ decays
with the junction length $L$ according to the power-law  $1/L$
(for $L\ll L_T$), rather than
exponentially.

The Josephson current in SNS junctions is affected strongly by the
quality of the interface.  The transmittance of interfaces between
semiconductors and superconductors varies widely, depending on the
nature of the junction. The interface that has been studied most
intensively in recent years, particularly in the context of mesoscopic
effects, is the Nb/InAs interface. This interface is unique in the
sense that a charge-accumulation layer is formed instead of a Schottky
barrier and, as a result, the interface transparency is quite high.
(According to a recent measurement of the proximity effect in this
structure, the interface transmission coefficient $T_0\approx 0.7$
\cite{REF:vanwees}.)\thinspace\ More commonly, however, the
transmittance may be quite low, both because of interface roughness and
Schottky-barriers formation.  Below, we calculate the Josephson current
through the LL in two limiting cases: perfectly transmitting interfaces
(Sec.~\ref{sec:hti}) and poorly transmitting interfaces
(Sec.~\ref{sec:pti}). The latter case has been investigated in
Ref.~\cite{REF:Hekking}.
\subsection{Perfectly transmitting interfaces}
\label{sec:hti}
First, we consider the case of perfectly transmitting interfaces,
in which the only scattering that takes place at the S/LL boundaries is
Andreev reflection, single-particle reflection being absent. The
Josephson current ${\cal J}$ is given by
\begin{equation}
\label{eq:jcdef}
{\cal J}=\frac{2e}{\hbar}\frac{\partial}{\partial \chi}\Omega,
\end{equation}
where $\Omega=-k_{B}T\ln Z$ is the grand po{\-}tential, and $Z$ is the grand
partition function.
Substituting
Eqs.~(\ref{eq:phirho},\ref{eq:phisig},\ref{eq:thetarho},\ref{eq:thetasig})
into Eq.~(\ref{eq:llham})
and diagonalizing the nonzero-mode part via a canonical transformation,
we get for the many-body eigenenergies
of the system
\begin{eqnarray}
{\cal E}&=&\frac{\pi \hbar}{4L}\left[v_{\rho}K_{\rho}\left(J'+
\frac{\chi}{\pi}\right)^2+\frac{v_{\sigma}}{K_{\sigma}}
M^2\right]
\nonumber\\
&&\qquad\qquad+\hbar \sum _{k>0}\sum_{\mu=\rho,\sigma}
v_{\mu}k(n_{k \mu}+1/2),
\label{eq:eigenst}\end{eqnarray}
where $n_{k \mu}^{}\equiv b_{k \mu}^{\dagger}b_{k \mu}^{}$, and
the new boson operators $b_{k \mu}$ are connected to
the old ones via
$a_{\mu}=b_{\mu}\cosh\lambda_{\mu} -\nolinebreak
b_{\mu}^{\dagger}\sinh\lambda_{\mu}$, in which
\begin{equation}
\label{eq:canon}
\lambda_{\rho/\sigma}=\pm\frac{1}{2}\tanh^{-1}\frac{1-K_{\rho/\sigma}^2}
{1+K_{\rho/\sigma}^2}.
\end{equation}
We see that the phase-difference $\chi$ appears only in the topological part
of ${\cal E}$, as it should, because the nonzero-mode excitations are neutral,
and therefore do not contribute to the (equilibrium) charge current.
 We also note that only two of the four
charge/spin bosons, viz., $\phi_{\sigma}$ and $\theta_{\rho}$,
contribute to the topological part of ${\cal E}$. (The Josephson-current
problem differs in this respect from the persistent-current problem, in
which all four bosons contain  topological excitations.)\thinspace\
The combination $\{\phi_{\sigma},\theta_{\rho}\}$ commonly arises  in the study
of
superconductivity in LLs \cite{REF:Emery}.

The partition function factorizes as $Z=Z_{\rm t}(\chi)Z_{\rm n}$,
where $Z_{\rm t/n}$ is the contribution from the topological
(nonzero-mode) part of ${\cal E}$. To calculate ${\cal J}$,
we need only know $Z_{\rm t}$, which is given by
\begin{equation}
\label{eq:zt1}
Z_{\rm t}(\chi)=
{\sum\limits_{J',M}}^{\prime}
e^{-\varepsilon_{\rho}
\left(J'+\chi/\pi\right)^2} e^{-\varepsilon_{\sigma}M^2},
\end{equation}
where $\varepsilon_{\rho}\equiv\pi L_Tv_{\rho}K_{\rho}/4v_{\rm F}L$
and $\varepsilon_{\sigma}\equiv\pi v_{\sigma}L_T/4K_{\sigma}v_{\rm F}L$,
and  the primed sum  indicates that $J'$ and $M$ are connected via the
constraint
found in Sec.~\ref{sec:bos} (i.e., $J'+M$ even).
(Although the spin-part of $Z_{\rm t}$ does not depend on $\chi$,
it does not simply reduce to an overall multiplicative factor because of this
constraint.)\thinspace\ It is convenient to represent the winding numbers
$J'$ and $M$ in the following form:
$J'=2j+\kappa_{\rm J}$ and
$M=2m+\kappa_{\rm M}$ (with $\kappa_{\rm J}=0,1$ and $\kappa_{\rm M}=0,1$)
\cite{REF:Loss}.
The topological constraint is then satisfied
for $j,m=0,\pm 1,\dots$ and
$\kappa_{\rm J}=\kappa_{\rm M}$.
We can then re-write Eq.~(\ref{eq:zt1}) in the unconstrained form:
\begin{equation}
\label{eq:zt2}
Z_{\rm t}(\chi)=
f_{0,\rho}(\chi)f_{0,\sigma}(0)+f_{1,\rho}(\chi)f_{1,\sigma}(0),
\end{equation}
where
\begin{equation}
\label{eq:sums}
f_{\kappa,\mu}(\chi)\equiv
\sum_{n=-\infty}^{\infty}e^{-\varepsilon_{\mu}\left(2n
+\kappa+\chi\right)^2}.
\end{equation}
The Josephson current can  now readily be calculated.  Without writing
down the exact expression (which contains, as usual for this kind of
problem, Jacobi $\vartheta_{3}$-functions \cite{REF:Loss}), we consider
only the asymptotic cases of low ($L\ll L_T$) and high
($L\gg L_T$) temperatures. In the former case, one finds
\begin{equation}
\label{eq:jlow}
{\cal J}=\frac{ev_{\rho}K_{\rho}}{L}\;\;\frac{\chi}{\pi},\;\;
\mbox{for}\;\;|\chi|\leq\pi,
\end{equation}
with ${\cal J}(\chi+2\pi)={\cal J}(\chi)$.
We note that the interaction-renormalization
 of the Josephson current is the same
as that of the persistent current \cite{REF:Loss}.
 When the Luttinger-liquid Hamiltonian (\ref{eq:llham}) is
obtained as the long-wavelength limit of the Hubbard Hamiltonian then
$v_{\rho}K_{\rho}=v_{\rm F}$.  Comparing Eq.~(\ref{eq:jlow}) for
$v_{\rho}K_{\rho}=v_{\rm F}$ with the
corresponding expressions for the non-interacting electrons
\cite{REF:Ishii,REF:Svid,REF:Bardeen}, we see that {\it the Josephson
current through the Luttinger liquid is precisely the same as through
the non-interacting electron gas}.  A word of caution is necessary,
however:  this conclusion is only valid if backscattering and Umklapp
scattering are not taken into account. Even if these types of
scattering are irrelevant (in the renormalization-group sense), they
will modify the parameters of the LL entering Eq.~(\ref{eq:llham}),
so that the equality $v_{\rho}K_{\rho}=v_{\rm F}$ will no longer hold
\cite{fn_irr}.
Nevertheless, the deviations from this equality are expected to
be small. (For instance, in the spinless case, the maximal
reduction in the product $v_{\rho}K_{\rho}$ due to Umklapp scattering
is 20\%, even at half-filling, when such processes are most effective
\cite{REF:Loss,REF:LM}).)\thinspace\ Also, there is a much more
significant source of the renormalization of ${\cal J}$ which we have
not yet taken into account, viz., the non-perfectness of the interfaces
(see Sec.~\ref{sec:pti}).

For high temperatures ($L\gg L_T$) we find
\begin{equation}
\label{eq:jhigh}
{\cal J}=8e\hbar^{-1}T
\exp\left(-2\pi\alpha L/L_T
\right)\sin\chi,
\end{equation}
where
\begin{equation}
\alpha\equiv
\frac{1}{2}\left(\frac{v_{\rho}K_{\rho}}{v_{\rm F}}
+\frac{v_{\sigma}}{v_{\rm F}K_{\sigma}}\right).
\label{eq:alpha}\end{equation}
In the Hubbard model with the SU(2) symmetry,
$\alpha=(1+1/\sqrt{1-g})/2>1$. Thus, at high
temperatures, interactions lead to the further suppression of ${\cal J}$
(in addition to the thermal disruption of the phase-coherence).
\subsection{Poorly transmitting interfaces}
\label{sec:pti}
Having discussed the case of perfectly transmitting interfaces, we now
give a brief discussion of the case of poorly-transmitting interfaces.
In this case, qualitative information can be obtained by making use of the
known results on the interaction-induced renormalizations of the
transmission coefficient. (For the analogous treatment of persistent
current in imperfect LL rings, cf.~Ref.~\cite{REF:GP}.)\thinspace\

First, consider a non-interacting SNS \lq\lq clean\rq\rq\ system (i.e.,
the elastic
mean free path being far greater than L), with interface transmission
coefficients $T_{1,2}\ll 1$.  For simplicity, we now restrict attention
to the low-temperature case ($L_T\gg L$). The result for
the Josephson current can be obtained from the general formula of
Ref.~\cite{REF:ALO}, Eq.~(16), which expresses $\cal J$ through the
probability for an excitation to propagate from one interface to
another within a certain time. Substituting the probability of
ballistic propagation into Eq.~(16) of Ref.~\cite{REF:ALO}, we find,
after some simple algebra, that the critical current ${\cal J}_c^{i}$
for the structure with imperfect interfaces is
\begin{equation}
{\cal J}_c^{i}\simeq T_1 T_2{\cal J}_c,
\label{eq:jcimp}\end{equation}
where ${\cal J}_c$ is the critical current for $T_1=T_2=1$, which is given
by the $\chi$-independent factor in Eq.~(\ref{eq:jlow}). In the
interacting case, the low-transparency interfaces can be described
within the weak-link model of Kane and Fisher \cite{REF:KF1,REF:KF2}. In this
model, the weak links (in our case, the interfaces) are treated as
perturbations that transfer  electrons between the disconnected (in zeroth
order) parts of the LL, and the renormalization group (RG) of the
boundary sine-Gordon model is used to find the effective values of the
hopping (transmission) probabilities.  The case of the double barrier
has been considered in Ref.~\cite{REF:KF2}. We note that: (i)~because in our
Hamiltonian (\ref{eq:llham}), and hence in our action, the topological
excitations are decoupled from the  nonzero-modes, the RG flow equations
for the transmission coefficients are the same as in Ref.~\cite{REF:KF2};
(ii)~the effective cut-off for the RG flow is provided in our case by
the junction length $L$. Therefore, we can borrow the result for the
renormalized product $T_1 T_2$ from the Kane-Fisher result for the
double barrier away from the resonance:
\begin{equation}
T_1T_2\rightarrow
T_1T_2(k_{\rm F}L)^{-\left(1/K_\rho+1/K_\sigma-2\right)}.
\label{eq:theft}\end{equation}
If the SU(2)
symmetry of the underlying Hubbard model is intact, i.e., $K_\sigma=1$,
we find
\begin{equation}
\label{eq:jcimpint}
{\cal J}_c^{i}
\simeq
T_1 T_2\left(\frac{1}{k_{\rm F}L}\right)^{K_\rho^{-1}-1}
{\cal J}_c,
\end{equation}
which is in agreement with Ref.~\cite{REF:Hekking},
up to a (non--universal) numerical coefficient.

Whether Eq.~(\ref{eq:jlow}) or Eq.~(\ref{eq:jcimpint}) is relevant to a
given experimental situation, depends on the bare (i.e.,
unrenormalized) values of $T_{1,2}$ and on $L$.  Suppose that
$T_{1}\approx T_{2}\equiv T_0\approx 1$. Then the interface barriers
can be treated according to the weak-barrier model
\cite{REF:KF1,REF:KF2}.  Assume, for simplicity, that the potential
barriers are $\delta$-functions with the (bare) amplitude $V_{0}$. As $V_{0}$
is small, its RG flow at distinct interfaces is independent, and given
by $V=V_0(L/a)^{(1-K_{\rho})/2}$.  Then $T_0$ is renormalized to
\begin{equation}
T=\frac{1}{1+(mV/\hbar
k_{\rm F})^2}=\frac{1}{1+\frac{1-T_0}{T_0}
\left(\frac{L}{a}\right)^{(1-K_{\rho})}}.
\end{equation}
For relatively short junctions, i.e.,
\begin{equation}
L\ll L^{*}
\simeq a\left(\frac{T_0}{1-T_0}\right)^{1/(1-K_\rho)},
\label{eq:dstar}
\end{equation}
the renormalization of $T_0$ due to interactions is small,
and Eq.~(\ref{eq:jlow}) applies. The better the interface the
larger $L^*$.  In particular, as $T_0\rightarrow 1$, $L^*\rightarrow\infty$,
in accordance with the previously-found virtual absence of the
renormalization of $\cal J$ for perfect interfaces (cf.~Sec.~\ref{sec:hti}).
For longer junctions, i.e., $L\gg L^*$, Eq.~(\ref{eq:jcimpint}) applies.

Choosing $g=Ua/\pi v_{\rm F}=1/2$ (i.e., $K_{\rho}\approx 0.8$), using
the value $T_{0}=0.7$ \cite{REF:vanwees} and recalling
that $a\simeq 1/k_{\rm F}\simeq 100$\AA\  in the relevant semiconductor
structures,
we find
$L^*\simeq 1\,\mu$m. Junctions of lengths in the range $0.1-10\,\mu$m are
quite common in experiments \cite{REF:exp.review.SSm}, so both
Eqs.~(\ref{eq:jlow}) and
(\ref{eq:jcimpint}) may be relevant in experimental situation.

Indirectly, one also can appreciate  the extent to which interactions
renormalize the Josephson current by using recent experimental results on
the (dissipative) conductance of the
ultra-high mobility GaAs quantum wires \cite{REF:NTT}. As was shown
in Ref.~\cite{REF:NTT}, the conductance of the wire is reduced from
the conductance quantum (i.e., $e^2/h$ per spin projection) as the temperature
is lowered, the temperature-dependence being consistent with the theory
of charge transport in dirty Luttinger liquids
 \cite{REF:Apel,REF:Fukuyama,REF:recent}.
The absolute value of this reduction is quite small, however: it amounts
to $1-5\%$ for wires of  length $2-10\,\mu$m.

\section{Proximity effect in Luttinger liquids}
\label{sec:prox}
As has been mentioned in Sec.~\ref{sec:aqp}, Andreev
reflection at the NS interface gives rise to
correlations
between electron- and hole-like excitations in N.
These correlations are similar to those between the single-particle
excitations in S, which can be viewed as the induction of superconducting
off-diagonal long-range order in N due to the proximity of S.
The presence of such order is usually
described by the (inhomogeneous) condensate wavefunction \cite{REF:Deutscher}
$F(x)$, defined by
\begin{equation}
F(x)\equiv\langle
\psi_{\uparrow}(x)\psi_{\downarrow}(x)\rangle
\label{eq:fdef}.
\end{equation}
In the bulk of N, $F=0$ . The scale over which $F$  (exponentially) changes
from its value
at the NS boundary to zero in the bulk is given by $L_T$. As $T\to 0$, the
length $L_T\to\infty$, and the exponential decay of $F$ crosses over to
 a slower
(power-law) decay. In particular, if N is a Fermi-liquid metal, $F$ decays with
the distance from the interface as $1/x$ (at $T=0$) \cite{REF:Falk}.
We now explore
how this decay law is changed if N is in  the LL  state.

Consider a semi-infinite LL occupying the half-line $x>\nolinebreak 0$ and
connected to
S at $x=0$.
The bosonized form of $F(x)$ is given by
\begin{equation}
\label{eq:Fbos}
F(x)=\frac{1}
{\pi\delta a}\Big\langle e^{-i\sqrt{2\pi}\theta_{\rho}(x,0)}
\cos\big(\sqrt{2\pi}
\phi_{\sigma}(x,0)\big)
\Big\rangle,\end{equation}
where $\delta\to +0$ is a (dimensionless) cut-off parameter, $a$ is the
microscopic scale of the system, and  $\theta_{\rho}(x,\tau)$
and $\phi_{\sigma}(x,\tau)$ are  boson fields in the (imaginary time)
Heisenberg representation.  In Eq.~(\ref{eq:Fbos}),
the average is taken with respect to Boltzmann factor $e^{-S/\hbar}$, where
$S=S_{\rho}+S_{\sigma}$
is the (Euclidean) action corresponding to Hamiltonian Eq.~(\ref{eq:llham}),
and
\begin{mathletters}
\begin{eqnarray}
S_{\rho} &\equiv& \frac{\hbar K_{\rho}}{2}\int dx d\tau \frac{1}{v_{\rho}}\left
(\partial_\tau\theta_{\rho}\right)^2+v_{\rho}
\left
(\partial_x\theta_{\rho}\right)^2,\label{eq:action_c}\\
S_{\sigma} &\equiv& \frac{\hbar}{2K_{\sigma}}
\int dx d\tau \frac{1}{v_{\sigma}}\left
(\partial_{\tau}\phi_{\sigma}\right)^2+v_{\sigma}
\left
(\partial_{x}\phi_{\sigma}\right)^2.
\label{eq:action_s}\end{eqnarray}\end{mathletters}(Note
 that we have deliberately expressed $S_{\mu}$ via those boson fields
that enter the bosonized form of $F(x)$.)\thinspace\ The presence of S at $x=0$
imposes certain
boundary conditions on these fields. We derive these boundary conditions
directly from the boundary conditions for fermions
Eq.~(\ref{eq:lrbc1a},\ref{eq:lrbc1b}) by using the bosonized form of the
fermion fields (\ref{eq:mand}). (The phase of the order parameter
in S  is now taken to be zero,
as we do not consider charge-flow through the
the interface.)\thinspace\  Simple algebra then leads to:
\begin{equation}
\label{eq:bcinf}
\phi_{\sigma}(0,\tau)=-\sqrt{2\pi}/4,\qquad \theta_{\rho}(0,\tau)=0.
\end{equation}
In the (semi)-infinite geometry, the energy of
topological excitations is infinitesimally small \cite{REF:fn.topol.en.}, and
therefore we do not have to
incorporate the winding numbers of such excitations in boundary conditions
(\ref{eq:bcinf}). As one might have anticipated, Andreev boundary conditions
for fermions (\ref{eq:lrbc1a},\ref{eq:lrbc1b}) impose boundary conditions only
on those components of the boson fields that occur in the bosonized form
of the condensate wavefunction (\ref{eq:Fbos}).

In order to remove the divergence in Eq.~(\ref{eq:Fbos})
as $\delta\rightarrow +0$,  we
use the following trick. Consider the modified boundary
condition for the $\phi_{\sigma}$-field:
$\phi_{\sigma}(0,\tau)=-\sqrt{2\pi}/4+\delta$. Introduce a new field
${\tilde\phi}_{\sigma}\equiv\sqrt{2\pi}/4-\delta+\phi_{\sigma}$ satisfying
the homogeneous boundary condition ${\tilde\phi}_{\sigma}(0,\tau)=0$.
After this, $F$ takes the form
\begin{eqnarray}
\label{eq:F.interm}
F(x) &=& \frac{\sin\delta}{\delta}\frac{1}{\pi a}
\Big\langle e^{-\sqrt{2\pi}\theta_\rho}\Big\rangle
\Big\langle
e^{-\sqrt{2\pi}{\tilde\phi}_\sigma}\Big\rangle
\Big|_{\delta\rightarrow 0}\nonumber\\
&=&
\frac{1}{\pi a}\exp\left\{-\pi\left(
G_{\rho}(x,x,0)+G_{\sigma}(x,x,0)\right)\right\},
\end{eqnarray}
where $G_{\rho/\sigma}(x,x',\tau)$ is the propagator of the charge
(spin) boson field, which satisfies
\begin{equation}
\label{eq:laplace}
K_{\mu}^{\alpha_{\mu}}(\partial_x^2+v_{\mu}^{-2}\partial_\tau^2)G_{\mu}=
-\delta(x-x')\delta(\tau),
\end{equation}
in which $\alpha_{\rho/\sigma}=\pm 1$. $G_\mu$ obeys the following boundary
conditions: $G_\mu (0,x',\tau)=0$, $G_{\mu}(x,x',\tau)|_{x\to \infty}
\rightarrow 0$ and $G_{\mu}(x,x',\tau+\beta)=G_{\mu}(x,x',\tau)$, where
$\beta=1/T$. The Fourier transform in $\tau$ of the solution of
Eq.~(\ref{eq:laplace}) is given by
\begin{equation}
\label{eq:G}
G_{\mu}(x,x',\omega)=K_{\mu}^{-\alpha_\mu}|{\bar \omega}|^{-1}
\sinh (|{\bar \omega}|x_{<})\exp(-|{\bar \omega}|x_{>}),
\end{equation}
where $x_{<}\equiv\mbox{min}\{x,x'\}$ and $x_{>}\equiv\mbox{max}\{x,x'\}$.
Inverting the transform, we get
\begin{equation}
\label{eq:asympt}
G_{\mu}(x,x,0)=\frac{1}{2\pi K^{\alpha_{\mu}}}\ln\left(x/a\right),
\end{equation}
where, in order to regularize $G_{\mu}$,
 we have chosen the same short-distance  cut-off $a$
 as in Eq.~(\ref{eq:Fbos}).
Substituting Eq.~(\ref{eq:asympt}) into Eq.~(\ref{eq:F.interm}), we find
\begin{equation}
\label{eq:scaling}
F(x)=\frac{C}{a}\left(\frac{a}{x}\right)^{\gamma},\;\;\mbox{with}\;\;
\gamma\equiv\frac{1}{2}(K_{\sigma}+K_{\rho}^{-1}),
\end{equation}
where $C$ is a (non-universal) numerical coefficient.
In the absence of interactions, $K_{\rho}=K_{\sigma}=1$ and we return
to the $1/x$ scaling. In the presence of repulsive (attractive)
interactions, $\gamma>1$ ($\gamma<1$), and
the condensate amplitude in the LL decays faster (slower) than in the FL.
This result is in accord with one's intuition: the repulsive (attractive)
Coulomb interaction
weakens (strengthens) the superconducting
 state induced in N by Andreev reflection.
The  exponent $\gamma$ is one half of  the
exponent determining the spatial decay
 of the (singlet) superconducting fluctuations
in the infinite LL \cite{REF:Emery}.

At the first sight, the result that the profile of the condensate wavefunction
 in the LL decays faster than in  the FL  seems to contradict
to the results of Sec.~\ref{sec:jos}, in  which it was found
that the junction-length
dependence of the critical current is the same in the LL and the FL. Indeed, it
seems
natural to connect the $1/x$ decay law of the condensate in the FL with
the $1/L$ dependence of ${\cal J}_c$; then, it would be reasonable to expect
that the
$1/x^\gamma$ decay law of the condensate in the LL would be
transformed into a $1/L^{\gamma}$-dependence of ${\cal J}_c$, even if
the interfaces are perfect \cite{fn_gamma}
 In fact, this conclusion
would not be valid and, as we show below, the $1/L$-dependence of ${\cal J}_c$
 (at $T=0$)
is universal, and not connected with the profile of the
condensate in the N region of an SNS junction. Consider, again, an SNS junction
of length $L\gg \xi_{\rm S}$. Our main argument is that at $T=0$
the only relevant lengthscale in the problem is $L$; therefore, at
distances from the interface larger than $\xi_{\rm S}$, the profile
 of condensate
wavefunction   N is described by a single dimensionless parameter
$x/L$. Therefore, $F(x)$ can be represented in the following form:
\begin{equation}
F(x)=F_0^{-}\Phi^{-}(x/L)+F_0^{+}\Phi^{+}(x/L),
\label{eq:single}
\end{equation}
where $F_{0}^{\pm}=F_0e^{i\chi_{1,2}}$ are the values of $F$ at
$x=0(L)$, and the scaling functions $\Phi^{\pm}(z)$
satisfy the following boundary conditions: $\Phi^{\pm}(1)=1 (0)$,
$\Phi^{\pm}(0)=0(1)$. The supercurrent flowing through the junction
is given by
\begin{equation}
\label{eq:current}
{\cal J}=iA\left(F(x)\frac{dF^{*}(x)}{dx}-\mbox{c.c.}\right),
\end{equation}
where $A$ is an $L$-independent constant. Substituting Eq.~(\ref{eq:single})
into Eq.~(\ref{eq:current}), we see that ${\cal J}$ can be represented in the
form
\begin{equation}
\label{eq:1/d}
{\cal J}=AL^{-1}\Phi(x/L),
\end{equation}
where $\Phi(z)$ is a scaling function that is a combination
of the functions $\Phi^{\pm}(z)$ and their derivatives.
 Due to charge conservation,
${\cal J}(x)$ does not depend on $x$, which can be satisfied only
if $\Phi(z)$ is $z$-independent.
Thus, we see that
 $J\propto 1/L$ regardless of
the particular form of the condensate wavefunction,
the latter determining only the numerical coefficient in front
of the $1/L$-dependence.
\section{Solution via the bosonization of the whole system}\
\label{sec:ms}
So far, we have applied bosonization only to the Luttinger-liquid part
of the system, i.e., to the interval $0<x<L$.  We can gain some
further insight into our results by comparison with a system in which the
LL occupies the entire real line, but, by some mechanism,
has acquired  a superconducting gap when $x<0$ and $x>L$
 (cf.~Fig.~\ref{FIG:fig1}c).  The
existence of a gap means that the usual Luttinger Hamiltonian is
modified by the addition of the term
\begin{equation}
H_{\rm gap}=\int dx |\Delta(x)|e^{i\chi(x)}
\left(
\psi^{\dagger}_{+\uparrow}\psi^{\dagger}_{-\downarrow}
+\psi^{\dagger}_{-\uparrow}\psi^{\dagger}_{+\downarrow}
\right)+\mbox{H.c}.
\label{EQ:ms_Xone}
\end{equation}
The corresponding  (Minkowski) bosonic action is
\begin{eqnarray}
S
&=&
\int dx dt
\Bigl\{
\frac{K_\rho}{2}
\left(
\frac 1{v_\rho}(\partial_t \theta_\rho)^2
-v_\rho(\partial_x \theta_\rho)^2
\right)
\nonumber
\\
&&
\qquad\qquad
+
\frac 1{2K_\sigma}
\left(
\frac 1{v_\sigma}(\partial_t
\phi_\sigma)^2-
v_\sigma(\partial_x \phi_\sigma)^2
\right)
\nonumber
\\
&&
\qquad\qquad
+
|\Delta|
\Big(
:\cos(\chi(x)+\sqrt{2\pi}
(\theta_\rho-\phi_\sigma)):
\nonumber
\\
&&
\qquad\qquad
+
:\cos(\chi(x)+\sqrt{2\pi}(\theta_\rho+\phi_\sigma)):
\Big)
\Bigr\},
\label{EQ:ms_Xtwo}
\end{eqnarray}where $\chi(x)$ is the local phase of the order parameter.
In regions where $\Delta(x)$ is large, the principal effect of the
non-linear terms is to constrain the values of $\theta_\rho(x)$ and
$\varphi_\sigma(x)$ to the minima of the cosine potential, so that
\begin{mathletters}
\begin{eqnarray}
\chi+\sqrt{2\pi}(\theta_\rho-\phi_\sigma)
&=&
2n\pi;
\label{EQ:ms_XthreeA}
\\
\chi+\sqrt{2\pi}(\theta_\rho+\phi_\sigma)
&=&
2m\pi.
\label{EQ:ms_XthreeB}
\end{eqnarray}
\end{mathletters}Equivalently
\begin{mathletters}
\begin{eqnarray}
\theta_\rho
&=&
\frac{1}{\sqrt{2\pi}}(-\chi+\pi(n+m));
\label{EQ:ms_XfourA}
\\
\phi_\sigma
&=&
\frac{1}{\sqrt{2\pi}}(+\pi(n-m)).
\label{EQ:ms_XfourB}
\end{eqnarray}
\end{mathletters}There are no constraints
 on $\theta_\sigma$ or $\phi_\rho$. The fields $\theta_{\rho}$ and
$\phi_{\sigma}$
are thus locked (modulo winding numbers) to the condensate phase in the
superconducting regions. In the purely LL part of the
system all four fields are free to fluctuate.  The condensate therefore
imposes boundary conditions that are essentially the same as those in
Eq.~(\ref{eq:bcinf}).

The bosonized form of the number density current is
\begin{equation}
j(x)=-2v_\rho K_\rho\frac
1{\sqrt{2\pi}}\partial_x\theta_\rho.
\label{EQ:ms_Xfive}
\end{equation}
Substituting $\theta_\rho$ from
Eq.~(\ref{EQ:ms_XfourA}) into Eq.~(\ref{EQ:ms_Xfive}) we find
the $T=0$ supercurrent to be
\begin{equation}
j(x)=2v_\rho K_\rho\frac{1}{2\pi}\partial_x \chi.
\label{EQ:ms_Xsix}
\end{equation}
We can confirm this result by considering the case of a Galilean-invariant
system. For such a system we know that
\begin{equation}
j(x)=\rho_s v_s=\rho_s\frac {\hbar}{2m}\partial_x \chi,
\label{EQ:ms_Xseven}
\end{equation}
where $\rho_s$ is the density of superconducting electrons.
At $T=0$ we will have  $\rho_s=\rho$.
Comparing Eqs.~(\ref{EQ:ms_Xsix}) and (\ref{EQ:ms_Xseven}),
we see that consistency requires the equilibrium number density in the
Galilean-invariant liquid
to be given by
\begin{equation}
\rho=2K_\rho v_\rho m/\pi\hbar.
\label{EQ:ms_Xeight}
\end{equation}
(The factor of $2$ in this equation arises from the two
spin-projections.)
That this is correct is shown  by comparing the commutator
\begin{equation}
[\psi^{\dagger}\psi(x), \frac{\hbar}{2mi}
\psi^{\dagger}(x'){\buildrel\leftrightarrow\over
\partial_{x'}}\psi(x')]=\frac{\hbar}{mi}
\psi^{\dagger}\psi(x)\partial_x \delta(x-x')
\label{EQ:ms_Xnine}
\end{equation}
of the charge and current in a Galilean invariant system with the
corresponding commutator in our Luttinger system, viz.,
\begin{equation}
[\rho(x), j(x')]=-2iK_\rho
v_\rho\frac{1}{\pi}\partial_x\delta(x-x').
\label{EQ:ms_Xten}
\end{equation}
The Luttinger model approximates the Galilean invariant system by the
replacement of the charge density operator on the right hand side of
Eq.~(\ref{EQ:ms_Xnine}) by its expectation value. This confirms that
Eq.~(\ref{EQ:ms_Xeight}) is correct.

We now apply Eq.(\ref{EQ:ms_Xseven}). In the purely LL segment of the line
(i.e., $0<x<L$)
the $\theta_\rho$ and $\phi_\sigma$ fields are no longer constrained by
the condensate. However, as we mentioned earlier, their values at the ends of
the interval are fixed, just as in Eq.~(\ref{eq:bcinf}):
\begin{equation}
\int_{x_1}^{x_2} j(x)dx=2v_\rho
K_\rho\frac{1}{2\pi}(\chi_2-\chi_1)
\label{EQ:ms_Xeleven}
\end{equation}
This is the same result as Eq.~(\ref{eq:jlow}),
because  the  quantity found by the
thermodynamic trick of differentiating the free energy with respect to
$\chi$ is the  spatial average of the current. The advantage of
Eq.~(\ref{EQ:ms_Xeight}) is that we can see that this average current is
independent of the precise way in which the gap goes to zero as we
enter the Luttinger link. Indeed, because the duality map between
one-dimensional  charge density waves (CDW)  and superconductors interchanges
the charge-
and current-densities, the results we have just described are just
the dual of the well-known result in the theory of
CDW systems that the total charge induced in a region is a
topological quantity depending only on the asymptotic values of the CDW
condensate phase \cite{REF:Gol_Wil}.
\section{Acknowledgements}
We thank Eduardo Fradkin for several useful discussions.
This work was supported by the US NSF under grants DMR89-20538 (DLM) and
DMR94-24511 (MS and PMG), and by the NSERC of Canada (DL).

\figure{(a)~A Luttinger liquid (LL) conductor connecting two
superconducting electrodes with phases of the order parameter
$\chi_1$ and $\chi_2$.
 (b)~The model profile of the pair-potential used for the derivation of Andreev
boundary conditions (Sec.~\ref{sec:abc}). (c)~Generic profile of the
pair-potential appropriate for the bosonization of the system as a whole
(Sec.~\ref{sec:ms}).\label{FIG:fig1}}

\begin{references}
\bibitem{REF:fn.normal} If N and S are metallurgically distinct, the
electrons arriving to the interface can also suffer normal reflection,
which would reduce the probability of Andreev reflection. This
issue will be dealt with in Sec.~\ref{sec:pti}.
\bibitem{fn_irr}The irrelevant interactions will give rise to the
corrections  in
the $L$- and $T$-dependences of ${\cal J}$, similar to those studied in
Ref.~\cite{REF:LM} for the case of persistent current in a  spinless
LL at half-filling.
\bibitem{REF:fn.topol.en.}As is seen from, e.g.,
Eqs.~(\ref{eq:mmin},\ref{eq:mplus}),
the energy of topological excitations in a finite-size system
decays with  system size as $\int_0^L dx
\left[\left(\partial \theta _\rho ^t\right )^2 +
\left(\partial \phi _\sigma ^t\right )^2\right]\propto L\times L^{-2}=1/L$,
where supercript $t$ denotes the topological-excitation parts of expansions
(\ref{eq:phirho})
and (\ref{eq:thetarho}).
\bibitem{fn_gamma} Note that the exponent $\gamma$ differs from the
exponents that determines the $L$-dependence of ${\cal J}$ for the case
of poorly transmitting interfaces [cf. Eq.~(\ref{eq:jcimpint})].
\end{references}
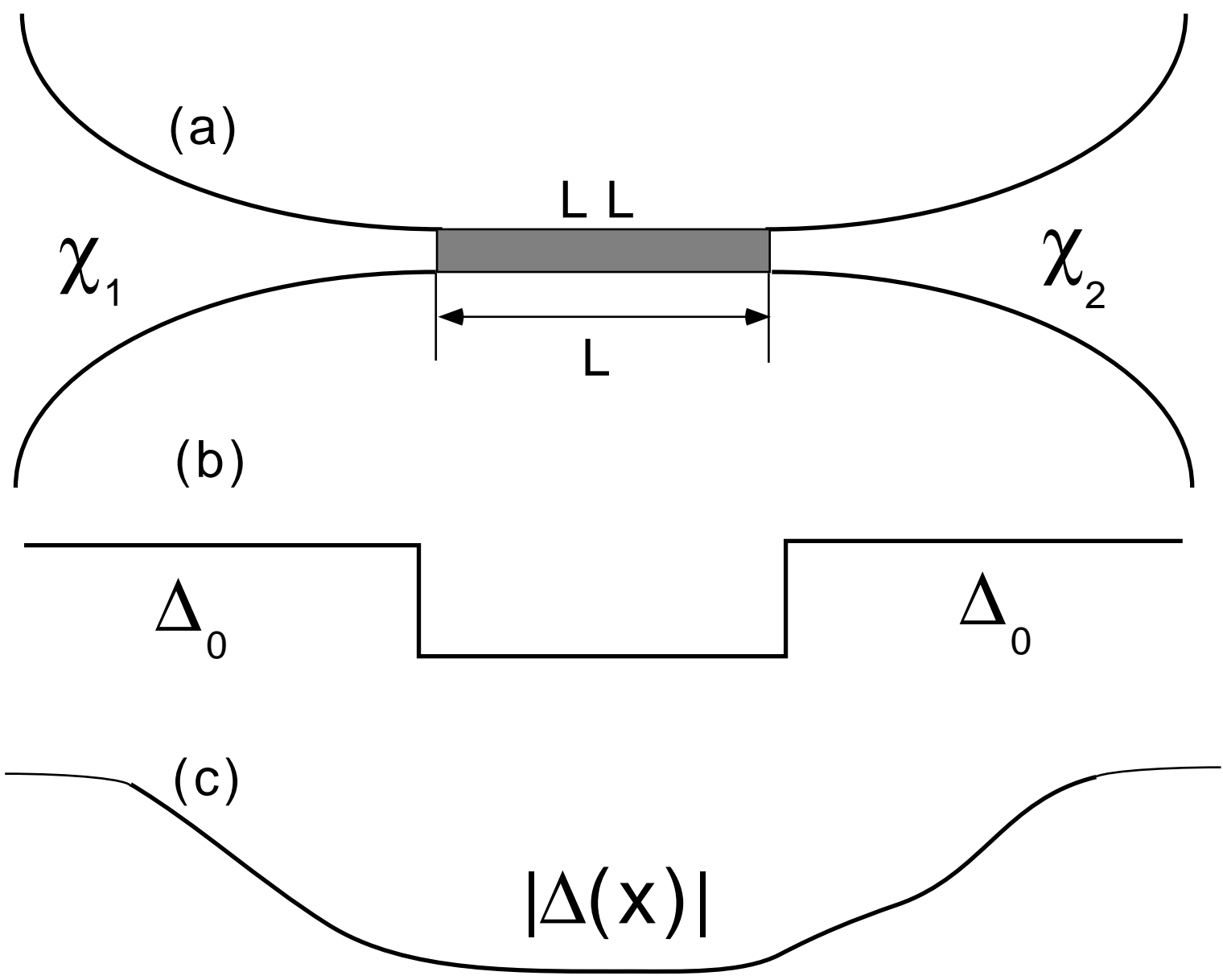
\end{document}